\documentclass[preprint,prd,aps,nofootinbib]{revtex4}
\usepackage{graphicx}
\usepackage{float}
\usepackage{diagbox}
\usepackage{booktabs}
\usepackage{color}
\usepackage{slashed}
\makeatletter

\newcommand{\Rmnum}[1]{\expandafter\@slowromancap\romannumeral #1@}
\makeatother
\begin{document}
\title{Kinetic energy and speed powers $v^n$ of a heavy quark inside $S$ wave and $P$ wave heavy-light mesons}
\author{Wei Li$^{1,2,3}$\footnote{watliwei@163.com},
Tianhong Wang$^{4}$\footnote{thwang@hit.edu.cn},
Tai-Fu Feng$^{1,2,3}$\footnote{fengtf@hbu.edu.cn},
Guo-Li Wang$^{1,2,3}$\footnote{wgl@hbu.edu.cn, Corresponding author},
Chao-Hsi Chang$^{5,6}$\footnote{zhangzx@itp.ac.cn, Corresponding author}}

\affiliation{&&${^1}$ Department of Physics, Hebei University, Baoding 071002, China
\nonumber\\$^{2}$ Hebei Key Laboratory of High-precision Computation and Application of Quantum Field Theory, Baoding 071002, China
\nonumber\\$^3$ Hebei Research Center of the Basic Discipline for Computational Physics, Baoding 071002, China
\nonumber\\$^4$ School of Physics, Harbin Institute of Technology, Harbin 150001, China
\nonumber\\$^5$ CAS Key Laboratory of Theoretical Physics, Institute of Theoretical Physics,
Chinese Academy of Sciences, Beijing 100190, China
\nonumber\\$^6$ School of Physical Sciences, University of
Chinese Academy of Sciences, Beijing 100049, China}
\begin{abstract}
Based on the instantaneous Bethe-Salpeter equation method, we calculate the average values $\overline{|\vec{q}|^n}\equiv q^n$ and speed powers $\overline{|\vec{v}|^n} \equiv v^n$ ($n=1,2,3,4$) of a heavy quark inside $S$ wave and $P$ wave heavy-light mesons, where $\vec{q}$ and $\vec{v}$ are the three dimensional momentum and velocity of the heavy quark, respectively. We obtain the kinetic energy $\mu^2_{_\pi}=0.455$ GeV$^2$ for the $B$ meson, which is consistent with the experimental result $0.464\pm 0.076$ GeV$^2$. For the $B_{s}$, $D$ and $D_{s}$ mesons, the $\mu^2_{_\pi}$ are $0.530$ GeV$^2$, $0.317$ GeV$^2$ and $0.369$ GeV$^2$, respectively. And $v^2=0.0185$, $0.0215$, $0.121$, and $0.140$ for $B$, $B_{s}$, $D$, and $D_{s}$. We obtain some relationships, for example, $q^n_{_{0^-}}(mS)\approx q^n_{_{1^-}}(mS)$,  $q^n_{_{0^+}}(mP)\approx q^n_{_{1^{+'}}}(mP^{'})> q^n_{_{1^+}}(mP)\approx q^n_{_{2^+}}(mP)$, and $q^n(mS)< q^n(mP)$ ($m=1,2,3$), etc.
\end{abstract}
\maketitle

\section{INTRODUCTION}

In the properties of heavy meson, the kinetic energy $\mu^2_{_\pi}$ (or $-\lambda_{_1}$) \cite{BigiI1994,FalkA1993,NeubertM2005} of a heavy quark inside a hadron is an important parameter for heavy quark effective theory (HQET) \cite{N.Isgur1989,N.Isgur1990,H.Georgi1991,M.Neubert1994}. When the contribution of the higher order term in the ${1}/{m_{_Q}}$ expansion is taken into account, the parameter $\mu^2_{_\pi}$ contributes to the inclusive semileptonic decays of the heavy hadron, so its value is important for precise calculations, and also affecting the determination Cabibbco-Kobayashi-Maskawa (CKM) matrix elements. Therefore, the kinetic energy has attracted considerable attention in recent years. In Ref. \cite{O.L.Buchm2006}, Buchmuller and Flacher obtained the result $\mu^2_{_\pi}= 0.401\pm0.040$ GeV$^2$ for $B$ meson. In non-relativistic calculations \cite{BriereRA2002,ChenS2001,CroninHennessyD2001}, the $\mu^2_{_\pi}$ values of pseudoscalar mesons and vector mesons are equal, including existing relativistic calculations \cite{HwangDS1997,FazioFD1996,KimCS2004}, which also ignore the $\mu^2_{_\pi}$ difference between pseudoscalar mesons and vector mesons, because the values of these HQET parameters are equal for pseudoscalar mesons and vector mesons in the leading order heavy quark expansion and nonrelativistic methods.

In addition, experimental evidences show that the dynamical scale of the heavy bound state is smaller than the heavy quark mass $m_{_Q}$. Consistent with this fact, the quark velocity $v$ is regarded as a small value in these systems, $v<1$ \cite{Caswell1986}. Thus, the dynamic scale of heavy meson can be hierarchically ordered, so that the corresponding amplitudes involving heavy meson can be systematically expanded in power of the quarks's typical velocity (as well as the momentum).

Nonrelativistic quantum chromodynamics (NRQCD) is a powerful and effective field theory to describe quark physics, under its framework, the typical velocity $v$ expansion method has been widely used, and the relevant review is in \cite{Bodwin1997}. In the perturbation region, some calculations also require $v^n$ values, for example in the framework of NRQCD, the expansion of $\alpha_{_s}$ is always accompanied by the expansion of $v$, $\alpha_{_s}(M)\sim v^2$ \cite{Bodwin1997}. This typical speed could be the expectation or the average value $\bar{v}$.
Without further calculation, the value $v$ or $v^n$ itself can be used to roughly estimate the relativistic correction is large or small. For example, Ref \cite{GeoffreyT2006} show us that, for $J/\psi$, $v^2\approx 0.3$, while $v^2\approx 0.1$ for $\Upsilon(1S)$, then we know that the relativistic correction of $J/\psi$ is larger than that of $\Upsilon$. In addition, $v^n$ is widely used when calculating the relativistic correction. For example, in the NRQCD method \cite{G.T.Bodwin2004,R.L.Zhu2017,F.Feng2017}, light-cone method \cite{W.Wang2017}, potential models \cite{D.Ebert2006,D.Ebert2009}, and lattice QCD \cite{G.T.Bodwin2002}, etc. Our previous works \cite{Z.K.Geng2019,G.L.Wang2020,W.Li2023} had found large relativistic corrections in quarkonia, especially in highly excited states. So precise $v^n$ values are more and more important in the physics of meson.

At present, some working groups have studied the typical speed of a heavy quark inside quarkonia by different methods. Such as, Gremm-Kapustin (GK) relation \cite{M.Gremm1997}, the equation of binding energy or kinetic energy \cite{R.L.Zhu2018}, potential models \cite{W.Buchmuller1981}, QCD sum rules \cite{V.V.Braguta2009}, light-front framework \cite{C.W.Hwang2009}, and extracting from experimental dada\cite{H.K.Guo2011}, etc. In our previous work \cite{GuoLiWang2020}, we also calculated the average values $q^n$ and $v^n$ of a heavy quark in quarkonia. However, the research on $q^n$ and $v^n$ of a heavy quark in a heavy-light meson, especially in a excited meson, is still relatively scarce. So this paper will calculate the values of $q^n$ and $v^n$ of a heavy quark in $S$ wave and $P$ wave heavy-light mesons, including highly excited states.

The Salpeter equation \cite{E.E. S1952} is instantaneous version of the BS equation \cite{E.E.SandH.A.B.1951}, it is a relativistic dynamic equation describing bound state. Applying the wave functions obtained by solving full Salpeter equation, we can get relatively accurate theoretical results, which are in good agreement with the experimental data \cite{C.Hsi.Chang2010,G.L.Wang2022,S.C.Li2018,T.h.Wang2017,Z.H.Wang2022,H.F.Fu2012}. So in this paper, the Salpeter equation method is chosen.

The remainder of this paper is organized as follows: In Sec. \Rmnum{2}, we give the wave functions and normalization conditions of various bound states. In Sec. \Rmnum{3}, we show how to calculate the average values $q^n$ and $v^n$. The numerical results and discussions are shown in Sec. \Rmnum{4}. The summary is in Sec. \Rmnum{5}. Introduction of the Bethe-Salpeter equation and Salpeter equation are presented in the Appendix.

\section{The wave functions and normalization conditions of $S$ wave and $P$ wave mesons}

In terms of the radial quantum number $n$, the spin $S$, the orbital angular momentum $L$ and the total angular momentum $J$, a meson can be tagged as $n^{2S+1}L_{_J}$. We can also label meson with $J^{P}$, where $P=(-1)^{L+1}$ is the parity. This representation of $n^{2S+1}L_{_J}$ holds in the nonrelativistic case, and $J^{P}$ holds in any case. In the absence of ambiguity, we still denote the particle state as $n^{2S+1}L_{_J}$ in this paper. In previous studies, the relativistic formula of the wave function for a meson with definite $J^{P}$ numbers is constructed requiring each term in the function having the same $J^{P}$ as the meson. With this wave function formula as input, the corresponding Salpeter equation is solved for different $J^{P}$ state, for example see Ref. \cite{C.Hsi.Chang2010}.

In this paper, for the heavy-light meson, we consider two $S$ wave states, pseudoscalar $^{1}S_{_0}$ ($0^{-}$) state and vector $^{3}S_{_1}$ ($1^{-}$) state, four $P$ wave states, $^{3}P_{_0}$ ($0^{+}$) state, $P_{_1}$ ($1^{+}$) state, $P^{'}_{_1}$ ($1^{+}$) state and $^{3}P_{_2}$ ($2^{+}$) state. Here we do not show the detail how to solve the corresponding Salpeter equation, only show the relativistic wave functions and normalization conditions of these states.

For the $S$ wave $0^{-}$ state, the relativistic wave function can be written as \cite{KimCS2004}
\begin{eqnarray}\label{0-}
&&\varphi_{_{0^{-}}}(q_{_\perp})=M\bigg[\gamma_{_0}a_{_1}+a_{_2}+\frac{\slashed{q}_{_\perp}}{M}a_{_3}
+\frac{\slashed{P}\slashed{q}_{_\perp}}{M^2}a_{_4}\bigg]\gamma_{_5},
\end{eqnarray}
where, $q_{_\perp}=(0,\vec{q})$, $M$ is the mass of the meson, $a_{_i}$ are the radial wave functions with respect to $-q^2_{_\perp}$. As can be seen from the last two equations in Eq. (\ref{SE2}), these radial wave functions $a_{_i}$ are not completely independent, and the constrained conditions between them are
\begin{eqnarray}\label{0-CC}
&&a_{_3}=\frac{a_{_2}M(\omega_{_{q}}-\omega_{_{Q}})}{(m_{_{Q}}\omega_{_{q}}+m_{_{q}}
\omega_{_{Q}})},
~~~~~~a_{_4}=-\frac{a_{_1}M(\omega_{_{q}}+\omega_{_{Q}})}{(m_{_{Q}}\omega_{_{q}}+m_{_{q}}
\omega_{_{Q}})}.
\end{eqnarray}
The normalization condition is
\begin{eqnarray}\label{n0-}
&&\int\frac{d^3\vec{q}}{(2\pi)^3}2a_{_1}a_{_2}M\bigg\{\frac{\omega_{_Q}-\omega_{_q}}{m_{_Q}-m_{_q}}+
\frac{m_{_Q}-m_{_q}}{\omega_{_Q}-\omega_{_q}}+\frac{2\vec{q}^2(\omega_{_Q}m_{_Q}+\omega_{_q}m_{_q})}{(\omega_{_Q}m_{_q}+\omega_{_q}m_{_Q})^2}\bigg\}=1,
\end{eqnarray}
where $\omega_{_Q}$, $\omega_{_q}$, $m_{_Q}$ and $m_{_q}$ are the energies and masses of heavy and light quarks, $\omega_{_{Q}}=\sqrt{m^2_{_{Q}}-q^2_{_{{\perp}}}}$ and $\omega_{_{q}}=\sqrt{m^2_{_{q}}-q^2_{_{{\perp}}}}$.

The relativistic wave function of $S$ wave $1^{-}$ state can be written as \cite{G.-L.Wang2006}
\begin{eqnarray}\label{1-}
&&\varphi_{_{1^{-}}}(q_{_\perp})=q_{_\perp}\cdot\epsilon^{\mu}_{_\perp}\bigg[b_{_1}+\frac{\slashed{P}}{M}b_{_2}
+\frac{\slashed{q}_{_\perp}}{M}b_{_3}+\frac{\slashed{P}\slashed{q}_{_\perp}}{M^2}b_{_4}\bigg]
+M\slashed{\epsilon}^{\mu}_{_\perp}b_{_5}+\slashed{\epsilon}^{\mu}_{_\perp}\slashed{P}b_{_6}+
\nonumber\\&&\hspace{1.9cm}(\slashed{q}_{_\perp}\slashed{\epsilon}^{\mu}_{_\perp}-q_{_\perp}\cdot\epsilon^{\mu}_{_\perp})b_{_7}
+\frac{1}{M}(\slashed{P}\slashed{\epsilon}^{\mu}_{_\perp}\slashed{q}_{_\perp}-\slashed{P}q_{_\perp}\cdot\epsilon^{\mu}_{_\perp})b_{_8},
\end{eqnarray}
where, $\epsilon^{\mu}_{_\perp}$ is the polarization vector of the $1^{-}$ state, radial wave functions are $b_{_i}=b_{_i}(-q^2_{_\perp})$. Four of the radial wave functions are independent, and we have the following relations
\begin{eqnarray}
&&b_{_1}=\frac{[q^2_{_\perp}b_{_3}+M^2b_{_5}](m_{_Q}m_{_q}-\omega_{_Q}\omega_{_q}+q^2_{_\perp})}{M(m_{_Q}+m_{_q})q^2_{_\perp}},
~~b_{_2}=\frac{[-q^2_{_\perp}b_{_4}+M^2b_{_6}](m_{_Q}\omega_{_q}-m_{_q}\omega_{_Q})}{M(m_{_Q}+m_{_q})q^2_{_\perp}},
\nonumber\\&&b_{_7}=\frac{M(-m_{_Q}m_{_q}+\omega_{_Q}\omega_{_q}+q^2_{_\perp})}{(m_{_Q}-m_{_q})q^2_{_\perp}}b{_{_5}},
~~~~~~~~~~~~b_{_8}=\frac{M(m_{_Q}\omega_{_q}-m_{_q}\omega_{_Q})}{(\omega_{_Q}-\omega_{_q})q^2_{_\perp}}b{_{_6}}.
\end{eqnarray}
The corresponding normalization condition is
\begin{eqnarray}\label{n1-}
&&\int\frac{d^3\vec{q}}{(2\pi)^3}\frac{8\omega_{_Q}\omega_{_q}}{3M}\bigg\{3b_{_5}b_{_6}\frac{M}{\omega_{_Q}m_{_q}+\omega_{_q}m_{_Q}}+
\frac{\omega_{_Q}\omega_{_q}-m_{_Q}m_{_q}+\vec{q}^2}{(m_{_Q}+m_{_q})(\omega_{_Q}+\omega_{_q})}\times
\nonumber\\&&\hspace{3cm}\bigg[b_{_4}b_{_5}-b_{_3}\bigg(b_{_4}\frac{\vec{q}^2}{M^2}+b_{_6}\bigg)\bigg]\bigg\}=1.
\end{eqnarray}

For the $P$ wave $0^{+}$ state, the relativistic wave function is written as \cite{G.-L.Wang2007}
\begin{eqnarray}\label{0+}
&&\varphi_{_{0^{+}}}(q_{_\perp})=\slashed{q}_{_\perp}c_{_1}+\frac{\slashed{P}\slashed{q}_{_\perp}}{M}c_{_2}
+Mc_{_3}+\slashed{P}_{_\perp}c_{_4},
\end{eqnarray}
radial wave functions $c_{_i}$ have the following relations
\begin{eqnarray}
&&c_{_3}=\frac{c_{_1}q^2_{_\perp}(m_{_{Q}}+m_{_{q}})}{M(\omega_{_{Q}}\omega_{_{q}}+m_{_{Q}}m_{_{q}}+q^2_{_\perp})},
~~~~~~c_{_4}=\frac{c_{_2}q^2_{_\perp}(\omega_{_{Q}}-\omega_{_{q}})}{M(m_{_{Q}}\omega_{_{q}}+m_{_{q}}\omega_{_{Q}})}.
\end{eqnarray}
The normalization condition is
\begin{eqnarray}\label{n0+}
&&\int\frac{d^3\vec{q}}{(2\pi)^3}\frac{8c_{_1}c_{_2}\omega_{_Q}\omega_{_q}\vec{q}^2}{M(\omega_{_Q}m_{_q}+\omega_{_q}m_{_Q})}=1.
\end{eqnarray}

For the two $P$ wave $1^{+}$ states, their formulas of wave functions are the same \cite{Q.Li2020}
\begin{eqnarray}\label{1+}
&&\varphi_{_{1^{+}}}(q_{_\perp})=(q_{_\perp}\cdot\epsilon)\bigg(d_{_1}+\frac{\slashed{P}}{M}d_{_2}+\frac{\slashed{q}_{_\perp}}{M}d_{_3}
+\frac{\slashed{P}\slashed{q}_{_\perp}}{M^2}d_{_4}\bigg)\gamma_{_5}+
\nonumber\\&&\hspace{1.8cm}+\frac{i\varepsilon_{_{\mu\nu\alpha\beta}}\gamma^{\mu}P^{\nu}q^{\alpha}_{_\perp}\epsilon^{\beta}}{M}
\bigg(d_{_5}+\frac{\slashed{P}}{M}d_{_6}+\frac{\slashed{q}_{\perp}}{M}d_{_7}
+\frac{\slashed{P}\slashed{q}_{\perp}}{M^2}d_{_8}\bigg),
\end{eqnarray}
where $\epsilon$ is the polarization vector, $\varepsilon_{_{\mu\nu\alpha\beta}}$ is the Levi-Civita symbol, $d_{_i}$ are radial wave functions, and we have
\begin{eqnarray}
&&d_{_1}=\frac{[q^2_{_\perp}d_{_3}+M^2d_{_5}](-m_{_Q}m_{_q}-\omega_{_Q}\omega_{_q}+q^2_{_\perp})}{M(m_{_Q}-m_{_q})q^2_{_\perp}},
~~d_{_2}=\frac{[-q^2_{_\perp}d_{_4}+M^2d_{_6}](m_{_Q}\omega_{_q}+m_{_q}\omega_{_Q})}{M(m_{_Q}+m_{_q})q^2_{_\perp}},
\nonumber\\&&d_{_7}=\frac{M(m_{_Q}m_{_q}+\omega_{_Q}\omega_{_q}+q^2_{_\perp})}{(m_{_Q}-m_{_q})q^2_{_\perp}}d{_{_5}},
~~~~~~~~~~~~~~~~~d_{_8}=\frac{M(m_{_Q}\omega_{_q}+m_{_q}\omega_{_Q})}{(\omega_{_Q}-\omega_{_q})q^2_{_\perp}}d{_{_6}}.
\end{eqnarray}
The normalization condition is
\begin{eqnarray}\label{n1+}
&&\int\frac{d^3\vec{q}}{(2\pi)^3}\frac{2(m_{_Q}\omega_{_q}+m_{_q}\omega_{_Q})\vec{q}^2}
{3M\omega_{_Q}\omega_{_q}}(d_{_3}d_{_4}-2d_{_5}d_{_6})=1.
\end{eqnarray}

For the $P$ wave $2^{+}$ state, the relativistic wave function can be written as \cite{G.-L.Wang2009}
\begin{eqnarray}\label{2+}
&&\varphi_{_{2^{+}}}(q_{_\perp})=\epsilon_{_{\mu\nu}}q^{\nu}_{_\perp}\bigg\{q^{\mu}_{_\perp}
\bigg[e_{_1}+\frac{\slashed{P}}{M}e_{_2}+\frac{\slashed{q}_{_\perp}}{M}e_{_3}
+\frac{\slashed{P}\slashed{q}_{_\perp}}{M^2}e_{_4}\bigg]
\nonumber\\&&\hspace{1.8cm}+\gamma^{\mu}\bigg[Me_{_5}+\slashed{P}e_{_6}+\slashed{q}_{_\perp}e_{_7}\bigg]
+\frac{i}{M}e_{_8}\varepsilon_{_{\mu\alpha\beta\gamma}}P^{\alpha}q^{\beta}_{_{\perp}}\gamma^{\gamma}\gamma_{_5}\bigg\},
\end{eqnarray}
here, $\epsilon_{_{\mu\nu}}$ is polarization tensor, $e_{_i}$ are radial wave functions, and constrained conditions between them are
\begin{eqnarray}
&&e_{_1}=\frac{[q^2_{_\perp}e_{_3}+M^2e_{_5}](\omega^2_{_Q}+\omega^2_{_q})-M^2e_{_5}(\omega^2_{_Q}-\omega^2_{_q})}{M(m_{_Q}\omega_{_q}+m_{_q}\omega_{_Q})},
~~~e_{_2}=\frac{[q^2_{_\perp}e_{_4}-M^2e_{_6}](\omega_{_Q}-\omega_{_q})}{M(m_{_Q}\omega_{_q}+m_{_q}\omega_{_q})},
\nonumber\\&&e_{_7}=\frac{M(\omega_{_Q}-\omega_{_q})}{m_{_Q}\omega_{_q}+m_{_q}\omega_{_Q}}e{_{_5}},
~~~~~~~~~~~~~~~~~~~~~~~~~~~~~~~~~~~~e_{_8}=\frac{M(m_{_Q}+\omega_{_q})}{m_{_Q}\omega_{_q}+m_{_q}\omega_{_Q}}e{_{_6}}.
\end{eqnarray}
The corresponding normalization condition is
\begin{eqnarray}\label{n2+}
&&\int\frac{d^3\vec{q}}{(2\pi)^3}\frac{8\omega_{_Q}\omega_{_q}\vec{q}^2}{15M(\omega_{_Q}m_{_q}+\omega_{_q}m_{_Q})}
\bigg\{e_{_5}e_{_6}M^2\bigg[5+\frac{(m_{_Q}+m_{_q})(\omega_{_Q}m_{_q}-\omega_{_q}m_{_Q})}{\omega_{_Q}\omega_{_q}(\omega_{_Q}+\omega_{_q})}\bigg]
\nonumber\\&&\hspace{5.5cm}+e_{_4}e_{_5}\vec{q}^2\bigg[2+\frac{(m_{_Q}+m_{_q})(\omega_{_Q}m_{_q}-\omega_{_q}m_{_Q})}{\omega_{_Q}\omega_{_q}(\omega_{_Q}+\omega_{_q})}\bigg]
\nonumber\\&&\hspace{5.5cm}-2\vec{q}^2e_{_3}(e_{_4}\frac{\vec{q}^2}{M^2}+e_{_6})\bigg\}=1.
\end{eqnarray}
With these wave functions, the Salpeter equation Eq. (\ref{SE2}) can be solved, and then the mass spectrum and wave function values of corresponding states can be obtained.

\section{Calculations of average $q^n$ and  $v^n$}

%In the phenomenological model of Ref. \cite{G.Altarelli1982}, the kinetic energy is derived from the Fermi motion of the heavy quark inside the hadron, one expects $-\lambda_{_1}\approx p^2_{_F}$, where $p^2_{_F}$ is the Fermi momentum.
Since $q^2$ is a Lorentz scalar, we can free to choose a special framework to calculate this parameter. For simplicity we choose the CMS of the heavy-light meson. In the CMS of the heavy-light meson, the heavy quark momentum $p_{_Q}$, the light antiquark momentum $p_{_q}$ and relative momentum between quarks $q$ have relationship $p_{_{Q_\perp}}=q_{_\perp}=-p_{_{q_\perp}}$, so $q$ is the quark's momentum. From the normalized conditional expressions of the different states above, their form can be summarized as $\int d\vec{q} f^2(-q^2_{_{\perp}})=1$, which means that probability of finding quarks in the whole momentum space is unit, and $f^2(-q^2_{_{\perp}})d\vec{q}$ is the possibility of quark momentum being $q_{_{\perp}}\rightarrow q_{_{\perp}}+dq_{_{\perp}}$. So as with the Maxwell velocity distribution method, we define the average value, $\langle q^n\rangle\equiv\overline{|\vec{q}|^n}$.

The average values $q^n$ of a heavy quark inside a $0^-$ heavy-light meson, namely $S$ wave state, in term of the BS method, is written as
\begin{eqnarray}\label{an0-}
\langle q^n\rangle_{0^{-}}=\int\frac{d^3\vec{q}}{(2\pi)^3}2a_{_1}a_{_2}|\vec{q}|^nM\bigg\{\frac{\omega_{_Q}-\omega_{_q}}{m_{_Q}-m_{_q}}+
\frac{m_{_Q}-m_{_q}}{\omega_{_Q}-\omega_{_q}}+\frac{2\vec{q}^2(\omega_{_Q}m_{_Q}+\omega_{_q}m_{_q})}{(\omega_{_Q}m_{_q}+\omega_{_q}m_{_Q})^2}\bigg\}.
\end{eqnarray}

The average average values $q^n$ of a heavy quark inside $1^-$ heavy-light meson, another $S$ wave state, is
\begin{eqnarray}\label{an1-}
&&\langle q^n\rangle_{1^{-}}=\int\frac{d^3\vec{q}}{(2\pi)^3}\frac{8\omega_{_Q}\omega_{_q}|\vec{q}|^n}{3M}\bigg\{3b_{_5}b_{_6}\frac{M}{\omega_{_Q}m_{_q}+\omega_{_q}m_{_Q}}+
\frac{\omega_{_Q}\omega_{_q}-m_{_Q}m_{_q}+\vec{q}^2}{(m_{_Q}+m_{_q})(\omega_{_Q}+\omega_{_q})}\times
\nonumber\\&&\hspace{5cm}\bigg[b_{_4}b_{_5}-b_{_3}\bigg(b_{_4}\frac{\vec{q}^2}{M^2}+b_{_6}\bigg)\bigg]\bigg\}.
\end{eqnarray}

Similarly, for $P$ wave states, $0^{+}$ state, $1^{+}$ state, and $2^{+}$ state, their average $q^n$ are expressed as
\begin{eqnarray}\label{an0+}
&&\langle q^n\rangle_{0^{+}}=\int\frac{d^3\vec{q}}{(2\pi)^3}\frac{8c_{_1}c_{_2}\omega_{_Q}\omega_{_q}|\vec{q}|^{2+n}}{M(\omega_{_Q}m_{_q}+\omega_{_q}m_{_Q})},
\end{eqnarray}
\begin{eqnarray}\label{an1+}
&&\langle q^n\rangle_{1^{+}}=\int\frac{d^3\vec{q}}{(2\pi)^3}\frac{2(m_{_Q}\omega_{_q}+m_{_q}\omega_{_Q})|\vec{q}|^{2+n}}
{3M\omega_{_Q}\omega_{_q}}(d_{_3}d_{_4}-2d_{_5}d_{_6}),
\end{eqnarray}
\begin{eqnarray}\label{an2+}
&&\langle q^n\rangle_{2^{+}}=\int\frac{d^3\vec{q}}{(2\pi)^3}\frac{8\omega_{_Q}\omega_{_q}|\vec{q}|^{2+n}}{15M(\omega_{_Q}m_{_q}+\omega_{_q}m_{_Q})}
\bigg\{e_{_5}e_{_6}M^2\bigg[5+\frac{(m_{_Q}+m_{_q})(\omega_{_Q}m_{_q}-\omega_{_q}m_{_Q})}{\omega_{_Q}\omega_{_q}(\omega_{_Q}+\omega_{_q})}\bigg]
\nonumber\\&&\hspace{7cm}+e_{_4}e_{_5}\vec{q}^2\bigg[2+\frac{(m_{_Q}+m_{_q})(\omega_{_Q}m_{_q}-\omega_{_q}m_{_Q})}{\omega_{_Q}\omega_{_q}(\omega_{_Q}+\omega_{_q})}\bigg]
\nonumber\\&&\hspace{7cm}-2\vec{q}^2e_{_3}(e_{_4}\frac{\vec{q}^2}{M^2}+e_{_6})\bigg\}.
\end{eqnarray}
Where, $|\vec{q}|$ is the absolute magnitude of momentum. In above, the method of calculating the average value of momentum $q$ obviously shows that average value is also the expectation value, so we have the ralation
\begin{eqnarray}\label{anv}
&&q^n\equiv\overline{|\vec{q}|^n}\equiv\langle q^n\rangle,~~~~~~v^n\equiv\overline{|\vec{v}|^n}\equiv\langle v^n\rangle,
\end{eqnarray}
where, $\vec{v}=\frac{\vec{q}}{m_{_Q}}$.

\section{Numerical results and discussions}

In our calculation, the model parameters as, $a=e=2.7183$, $\alpha=0.06~\rm{GeV}$, $\Lambda_{_{QCD}}=0.21~\rm{GeV}$, in addition, $\lambda=0.24~\rm{GeV^2}$. The constituent quark masses, $m_{u}=m_{d}=0.38~\rm{GeV}$, $m_{s}=0.55~\rm{GeV}$, $m_{c}=1.62~\rm{GeV}$ and $m_{b}=4.96~\rm{GeV}$. Where we choose the Cornell potential \cite{E.Eichten1978}, a linear scalar potential plus a Coulomb vector potential. Since the predicted mass spectrum may not match very well with the experiment data, a free constant parameter $V_{0}$ is usually added to linear scalar potential to fit data \cite{S.G1985}. So by varying the $V_0$ \cite{C.Hsi.Chang2010}, we fit the experimental meson masses, and obtain the numerical values of the corresponding wave functions. For the masses of the ground state and the determined excited state, we refer to the values of PDG \cite{K.A.Olive2022}, and for the masses of the undetermined excited states, their values come from the results of the calculation by the BS equation method. In addition, we taken meson mass $M_{c\bar{u}}=M_{c\bar{d}}$, $M_{b\bar{u}}=M_{b\bar{d}}$, for example, $M_{D^0}=M_{D^\pm}=1864.84~\rm{MeV}$, $M_{B^0}=M_{B^\pm}=5279.66~\rm{MeV}$.

Using the Eqs. (\ref{an0-})-(\ref{an2+}), the average values of $q^n$ and $v^n$ for a heavy quark inside different heavy-light meson are calculated, and results are shown in Tables \ref{I}-\ref{X}. In these tables, we also show the masses of ground state mesons and excited states which used ground masses as inputs to solve the relativistic wave functions, then obtain the masses.

\subsection{The average values $q^n$ and $v^n$ of a bottom quark inside the $B$ and $B_{s}$ mesons}

The average values of $q^n$ of bottom quark inside $0^{-}$ state for the $B$, $B_{s}$ mesons and their excited states are shown in Table \ref{I}, their masses are also listed in table. According to the classification of $n^{2S+1}L_{_J}$, pseudoscalar $0^{-}$($^1S_{_0}$) state is $S$ wave state.

From Table \ref{I}, we can see that the kinetic energy $q^2$ of the $B$ meson is $0.455$ GeV$^2$, which is consistent with the latest result $\mu^2_{_\pi}=0.464\pm 0.076$ GeV$^2$ by the HFLAV collaboration \cite{Yasmine2023}, the experimental value $\mu^2_{_\pi}=0.434\pm 0.043$ GeV$^2$ given by the Belle collaboration \cite{Belle2021}, and $\mu^2_{_\pi}=0.460\pm 0.044$ GeV$^2$ by the Belle II collaboration \cite{Belle-II2022}. In the early days, a model-independent lower bound was established, $\mu^2_{_\pi}>0.4$ GeV$^2$ \cite{I.I.Bigi1995,M.Shifman1995,M.Voloshin1995}. Meanwhile, Bagan  predicted $\mu^2_{_\pi}=0.5\pm 0.15$ GeV$^2$ with the QCD sum rules \cite{E.Bagan1996}. Patricia also used the QCD sum rules to obtained $\mu^2_{_\pi}=0.54\pm 0.12$ GeV$^2$, and take into account the radiative corrections, $\mu^2_{_\pi}=0.60\pm 0.1$ GeV$^2$ \cite{Patricia1994}. Nikolai assumed $\mu^2_{_G}\approx0.4$ GeV$^2$, then contained $\mu^2_{_\pi}=0.5$ GeV$^2$ \cite{Nikolai1998}. Fulvia used the QCD relativistic potential model and obtained $\mu^2_{_\pi}=0.66$ GeV$^2$ \cite{FazioFD1996}. Hwang used the virial theorem obtained $\mu^2_{_\pi}=0.40\sim0.58$ GeV$^2$ \cite{HwangDS1997}. Gambino used value $0.40$ GeV$^2$ of $\mu^2_{_\pi}$ in job \cite{P.Gambino2011}, and Alberti taken the value of $\mu^2_{_\pi}$ to be $0.465\pm 0.068$ GeV$^2$  \cite{A.Alberti2015}. The value of $\mu^2_{_\pi}$ adopted by Bordone for the purpose of discussing the relation between pole and kinetic heavy quark masses is $0.477\pm 0.056$ GeV$^2$ \cite{M.Bordone2021}. Therefore, $\mu^2_{_\pi}$ has been calculated in various phenomenological models, and the theoretical results of $\mu^2_{_\pi}$ for the $B$ meson depend on models strongly.
\begin{table}
\caption{ The average values of $q^n$ and $v^n$ of the bottom quark inside the $0^{-}$ state $B$, $B_{s}$ and their excited states.}%, where the masses of $M_{B^0,\bar{B}^0,B^\pm}=5279.66~\rm{MeV}$, and $M_{B^0_s,\overline{B}^0_s}=5366.92~\rm{MeV}$ are input.}
\begin{tabular}{cccccccccc}
\hline
\hline
State $(B)$~~&~~{Mass}~~&~~{$q$}~~&~~{$q^2$}~~&~~{$q^3$}~~&~~{$q^4$}~~&~~~{$v$}~~&~~{$v^2$}~~&~~{$v^3$}~~&~~{$v^4$}~~\\
\hline
\hline
$1^1S_{_0}$~~&~~$5279.66$~~&~~$0.608$~~&~~$0.455$~~&~~$0.416$~~&~~$0.469$~~&~~$0.123$~~&~~$0.0185$~~&~~$0.00341$~~&~~$0.000775$~~\\
\hline
$2^1S_{_0}$~~&~~$5915.41$~~&~~$0.782$~~&~~$0.800$~~&~~$0.955$~~&~~$1.28$~~&~~$0.158$~~&~~$0.0325$~~&~~$0.00783$~~&~~$0.00211$~~\\
 \hline
$3^1S_{_0}$~~&~~$6315.78$~~&~~$0.902$~~&~~$1.04$~~&~~$1.37$~~&~~$2.07$~~&~~$0.182$~~&~~$0.0420$~~&~~$0.0112$~~&~~$0.00332$~~\\
 \hline
 \hline
State $(B_{s})$~~&~~{Mass}~~&~~{$q$}~~&~~{$q^2$}~~&~~{$q^3$}~~&~~{$q^4$}~~&~~~{$v$}~~&~~{$v^2$}~~&~~{$v^3$}~~&~~{$v^4$}~~\\
\hline
\hline
$1^1S_{_0}$~~&~~$5366.92$~~&~~$0.655$~~&~~$0.530$~~&~~$0.522$~~&~~$0.631$~~&~~$0.132$~~&~~$0.0215$~~&~~$0.00428$~~&~~$0.00104$~~\\
\hline
$2^1S_{_0}$~~&~~$5995.38$~~&~~$0.812$~~&~~$0.876$~~&~~$1.11$~~&~~$1.57$~~&~~$0.164$~~&~~$0.0356$~~&~~$0.00906$~~&~~$0.00259$~~\\
 \hline
$3^1S_{_0}$~~&~~$6385.15$~~&~~$0.924$~~&~~$1.12$~~&~~$1.56$~~&~~$2.42$~~&~~$0.186$~~&~~$0.0454$~~&~~$0.0128$~~&~~$0.00399$~~\\
\hline
\hline
\end{tabular}
\label{I}
\end{table}

We also calculate the $\mu^2_{_\pi}$ (is $q^2$ in this paper) of the excited states for the $B$ and $B_{s}$ mesons. For example, see Table \ref{I}, $q^2_{_B}(2S,3S)= \{0.800,1.04\}$ GeV$^2$ and $q^2_{_{B_s}}(1S,2S,3S)= \{0.530,0.874,1.11\}$ GeV$^2$. It can be seen that although it is the same heavy quark, its average kinetic energy varies greatly in different mesons or excited states. For example, the value $q^2=0.530$ GeV$^2$ of the $b$ quark in the $B_{s}$ meson is significantly larger than that of $q^2=0.455$ GeV$^2$ in the $B$ meson.
%The difference of $0.07$ GeV$^2$ is not a negligible value compared to the value $q^2$ itself. This means that the same bottom quark in meson is more constrained by a light partner quark than by a heavy one.
In addition, the $q^2$ in highly excited state is larger than in the ground state or lowly excited state.

The average value of $v^n$ is also widely used when calculating the relativistic corrections and $\alpha_s$, etc. Therefore, we calculated the average values $v^n$ of a bottom quark inside $B$, $B_{s}$ mesons and their excited states, and show the results in Table \ref{I}. Our results indicate that the average value $v^n$ in a highly excited state is larger than in a lowly excited state or the ground state. For example, in case of $\{1^1S_{_0},~2^1S_{_0},~3^1S_{_0}\}$ for $B$ meson, we obtain $v_{_{0^-}}=\{0.123,~0.158,~0.182\}$, $v^2_{_{0^-}}=\{0.0185,~0.0325,~0.0420\}$, $v^3_{_{0^-}}=\{0.00341,~0.0783,~0.0112\}$ and $v^4_{_{0^-}}=\{0.000775,~0.00211,~0.00332\}$. So there is the relation $v^n_{_{0^-}}(1S)<v_{_{0^-}}^n(2S)<v^n_{_{0^-}}(3S)$ ($n=1,2,3,4$). We believe that the reason of relation is that there are more nodes in the wave function of the highly excited state than those of the lowly excited state and the ground state. The node structure results in big contribution from large $v$ region, so we obtained big average values $\overline{v^n}$, which indicate very large relativistic corrections in highly excited state. For the $B_{_s}$ meson, we have the similar relation and conclusion.

\begin{table}[H]
\caption{ The average values of $q^n$ and $v^n$ of the bottom quark inside the $1^{-}$ state $B^*$, $B^*_{s}$ and their excited states.}%, where the masses $M_{B^*}=5324.71~\rm{MeV}$ and $M_{B^*_s}=5415.40~\rm{MeV}$ are input.}
\begin{tabular}{cccccccccc}
\hline
\hline
State $(B^*)$~~&~~{Mass}~~&~~{$q$}~~&~~{$q^2$}~~&~~{$q^3$}~~&~~{$q^4$}~~&~~~{$v$}~~&~~{$v^2$}~~&~~{$v^3$}~~&~~{$v^4$}~~\\
\hline
\hline
$1^3S_{_1}$~~&~~$5324.71$~~&~~$0.607$~~&~~$0.450$~~&~~$0.404$~~&~~$0.442$~~&~~$0.122$~~&~~$0.0183$~~&~~$0.00331$~~&~~$0.000730$~~\\
\hline
$2^3S_{_1}$~~&~~$5952.23$~~&~~$0.781$~~&~~$0.796$~~&~~$0.939$~~&~~$1.23$~~&~~$0.157$~~&~~$0.0324$~~&~~$0.00770$~~&~~$0.00204$~~\\
 \hline
$3^3S_{_1}$~~&~~$6363.05$~~&~~$0.896$~~&~~$1.03$~~&~~$1.36$~~&~~$2.00$~~&~~$0.181$~~&~~$0.0417$~~&~~$0.0111$~~&~~$0.00330$~~\\
 \hline
 \hline
State $(B^*_{s})$~~&~~{Mass}~~&~~{$q$}~~&~~{$q^2$}~~&~~{$q^3$}~~&~~{$q^4$}~~&~~~{$v$}~~&~~{$v^2$}~~&~~{$v^3$}~~&~~{$v^4$}~~\\
\hline
\hline
$1^3S_{_1}$~~&~~$5415.40$~~&~~$0.653$~~&~~$0.523$~~&~~$0.506$~~&~~$0.592$~~&~~$0.131$~~&~~$0.0231$~~&~~$0.00415$~~&~~$0.000979$~~\\
\hline
$2^3S_{_1}$~~&~~$6035.40$~~&~~$0.807$~~&~~$0.873$~~&~~$1.09$~~&~~$1.52$~~&~~$0.163$~~&~~$0.0355$~~&~~$0.00895$~~&~~$0.00251$~~\\
 \hline
$3^3S_{_1}$~~&~~$6444.40$~~&~~$0.917$~~&~~$1.11$~~&~~$1.55$~~&~~$2.41$~~&~~$0.185$~~&~~$0.0450$~~&~~$0.0127$~~&~~$0.00398$~~\\
\hline
\hline
\end{tabular}
\label{II}
\end{table}

The average values $q^n$ and $v^n$ of the bottom quark inside the $1^-$ states $B^*$, $B^*_{s}$ and their excited states are shown in Tables \ref{II}. Where we can see,  $q^n_{_{B^*}}(1S)< q^n_{_{B^*}}(2S)< q^n_{_{B^*}}(3S)$, $q^n_{_{B^*_s}}(1S)< q^n_{_{B^*_s}}(2S)< q^n_{_{B^*_s}}(3S)$, and $q^n_{_{B^*}}(mS)< q^n_{_{B^*_s}}(mS)$, ($n=1,2,3,4$; $m=1,2,3$). For $v^n$, there have similar relations to $0^-$ state. We also note that the $q^n$ and $v^n$ values inside a $1^-$ meson are not much different from the cases of $0^-$ meson, for example, $q^2_{_{B}}(1S)=0.455$ GeV$^2 \approx q^2_{_{B^*}}(1S)=0.450$ GeV$^2$, so we have the relations $q^n_{_{0^-}}(mS)\approx q^n_{_{1^-}}(mS)$ and $v^n_{_{0^-}}(mS)\approx v^n_{_{1^-}}(mS)$ for $B$ or $B_{s}$ mesons. This result is similar to the relationship in \cite{Bodwin1997} with respect to quarkonia and they are treated $v^n$ of $J/\psi$ and $\eta_{_c}$ or $\Upsilon$ and $\eta_{_b}$ as same value.

\begin{table}
\caption{ The average values of $q^n$ and $v^n$ of bottom quark inside $0^{+}$ state $B^*_0$, $B^*_{s0}$ and their excited states.}
\begin{tabular}{cccccccccc}
\hline
\hline
State $(B^*_0)$~~&~~{Mass}~~&~~{$q$}~~&~~{$q^2$}~~&~~{$q^3$}~~&~~{$q^4$}~~&~~~{$v$}~~&~~{$v^2$}~~&~~{$v^3$}~~&~~{$v^4$}~~\\
\hline
\hline
$1^3P_{_0}$~~&~~$5698.87$~~&~~$0.761$~~&~~$0.655$~~&~~$0.637$~~&~~$0.712$~~&~~$0.153$~~&~~$0.0266$~~&~~$0.00522$~~&~~$0.00117$~~\\
\hline
$2^3P_{_0}$~~&~~$6199.12$~~&~~$0.883$~~&~~$0.951$~~&~~$1.17$~~&~~$1.61$~~&~~$0.178$~~&~~$0.0387$~~&~~$0.00963$~~&~~$0.00265$~~\\
 \hline
$3^3P_{_0}$~~&~~$6510.41$~~&~~$0.972$~~&~~$1.15$~~&~~$1.57$~~&~~$2.35$~~&~~$0.196$~~&~~$0.0469$~~&~~$0.0128$~~&~~$0.00388$~~\\
\hline
\hline
State $(B^*_{s0})$~~&~~{Mass}~~&~~{$q$}~~&~~{$q^2$}~~&~~{$q^3$}~~&~~{$q^4$}~~&~~~{$v$}~~&~~{$v^2$}~~&~~{$v^3$}~~&~~{$v^4$}~~\\
\hline
\hline
$1^3P_{_0}$~~&~~$5828.76$~~&~~$0.817$~~&~~$0.753$~~&~~$0.787$~~&~~$0.935$~~&~~$0.165$~~&~~$0.0307$~~&~~$0.00645$~~&~~$0.00154$~~\\
\hline
$2^3P_{_0}$~~&~~$627.22$~~&~~$0.910$~~&~~$1.03$~~&~~$1.34$~~&~~$1.92$~~&~~$0.183$~~&~~$0.0418$~~&~~$0.0109$~~&~~$0.00317$~~\\
 \hline
$3^3P_{_0}$~~&~~$6569.95$~~&~~$0.991$~~&~~$1.23$~~&~~$1.76$~~&~~$2.77$~~&~~$0.200$~~&~~$0.0501$~~&~~$0.0144$~~&~~$0.00457$~~\\
\hline
\hline
\end{tabular}
\label{III}
\end{table}

The average values $q^n$ and $v^n$ of a bottom quark inside $0^+$, $1^+$($P_{_1}$, $P^{'}_{_1}$) and $2^+$ $B_J$ and $B_{sJ}$ mesons and their excited states are shown in Tables \ref{III}, \ref{IV} and \ref{V}, respectively. According to the classification of $n^{2S+1}L_{_J}$, they are all $P$ wave states. For these $P$ wave $B_J(nP)$ and $B_{sJ}(nP)$ ($J=0,1,2$), we have the relations, $q^n_{_{B_J}}(1P)< q^n_{_{B_J}}(2P)< q^n_{_{B_J}}(3P)$, $q^n_{_{B_{sJ}}}(1P)< q^n_{_{B_{sJ}}}(2P)< q^n_{_{B_{sJ}}}(3P)$, and $q^n_{_{B_J}}(mP)< q^n_{_{B_{sJ}}}(mP)$ ($n=1,2,3,4$; $m=1,2,3$). We also notice that $q^n_{_{0^+}}(mP)\approx q^n_{_{1^{+'}}}(mP^{'})> q^n_{_{1^+}}(mP)\approx q^n_{_{2^+}}(mP)$ for $B_J$ or $B_{sJ}$.
Compared with the results of $S$ waves, we find there are the following relations, $q^n_{_{B_J}}(mS)< q^n_{_{B_J}}(mP)$ and $q^n_{_{B_{sJ}}}(mS)< q^n_{_{B_{sJ}}}(mP)$. For example, take the $B^{(*)}_{(J)}$ mesons as examples, the kinetic energies of $S$ wave pseudoscalar and vector are almost equal, $q^2_{_{0^-}}(1S)\simeq q^2_{_{1^-}}(1S)\simeq 0.45$ GeV$^2$, but significantly smaller than that of $P$ wave. For $P$ wave $0^+$ and $1^{+'}$, they have similar values $q^2_{_{0^+}}(1P)\simeq q^2_{_{1^{+'}}}(1P)\simeq 0.66$ GeV$^2$, while they are slightly larger than those of $1^{+}$ and $2^+$, $q^2_{_{1^+}}(1P)\simeq q^2_{_{2^{+}}}(1P)\simeq 0.59$ GeV$^2$.

\begin{table}
\caption{ The average values of $q^n$ and $v^n$ of bottom quark inside $1^{+}$ state $B_1$ ($P_{_1}$), $B^{'}_1$ ($P^{'}_{_1}$), $B_{s1}$ ($P_{_1}$), $B^{'}_{s1}$ ($P^{'}_{_1}$) and their excited states.}%, where the masses $M_{B_{_1}(5721)^0}=5725.90~\rm{MeV}$, $M_{B_{_{s1}}(5830)^0}=5828.70~\rm{MeV}$ are input.}
\begin{tabular}{cccccccccc}
\hline
\hline
State $(B^{(')}_1)$~~&~~{Mass}~~&~~{$q$}~~&~~{$q^2$}~~&~~{$q^3$}~~&~~{$q^4$}~~&~~~{$v$}~~&~~{$v^2$}~~&~~{$v^3$}~~&~~{$v^4$}~\\
\hline
\hline
$1P_{_1}$~~&~~$5725.46$~~&~~$0.725$~~&~~$0.586$~~&~~$0.525$~~&~~$0.521$~~&~~$0.146$~~&~~$0.0238$~~&~~$0.00430$~~&~~$0.000861$~\\
\hline
$2P_{_1}$~~&~~$6194.74$~~&~~$0.868$~~&~~$0.905$~~&~~$1.07$~~&~~$1.38$~~&~~$0.175$~~&~~$0.0368$~~&~~$0.00874$~~&~~$0.00228$~\\
 \hline
$3P_{_1}$~~&~~$6499.32$~~&~~$0.974$~~&~~$1.18$~~&~~$1.712$~~&~~$3.01$~~&~~$0.196$~~&~~$0.0480$~~&~~$0.0140$~~&~~$0.00497$~\\
\hline
$1P^{'}_{_1}$~~&~~$5762.88$~~&~~$0.769$~~&~~$0.664$~~&~~$0.645$~~&~~$0.715$~~&~~$0.155$~~&~~$0.0270$~~&~~$0.00529$~~&~~$0.00118$~\\
\hline
$2P^{'}_{_1}$~~&~~$6224.28$~~&~~$0.884$~~&~~$0.956$~~&~~$1.20$~~&~~$1.750$~~&~~$0.178$~~&~~$0.0389$~~&~~$0.00986$~~&~~$0.00289$~\\
 \hline
$3P^{'}_{_1}$~~&~~$6518.84$~~&~~$0.983$~~&~~$1.21$~~&~~$1.83$~~&~~$2.45$~~&~~$0.198$~~&~~$0.0496$~~&~~$0.0150$~~&~~$0.00405$~\\
\hline
\hline
State $(B^{(')}_{s1})$~~&~~{Mass}~~&~~{$q$}~~&~~{$q^2$}~~&~~{$q^3$}~~&~~{$q^4$}~~&~~~{$v$}~~&~~{$v^2$}~~&~~{$v^3$}~~&~~{$v^4$}~\\
\hline
\hline
$1P_{_1}$~~&~~$5828.41$~~&~~$0.774$~~&~~$0.670$~~&~~$0.645$~~&~~$0.688$~~&~~$0.156$~~&~~$0.0272$~~&~~$0.00528$~~&~~$0.00114$~\\
\hline
$2P_{_1}$~~&~~$6275.45$~~&~~$0.898$~~&~~$0.986$~~&~~$1.232$~~&~~$1.70$~~&~~$0.181$~~&~~$0.0401$~~&~~$0.0101$~~&~~$0.00280$~\\
 \hline
$3P_{_1}$~~&~~$6674.23$~~&~~$1.01$~~&~~$1.27$~~&~~$1.79$~~&~~$2.76$~~&~~$0.204$~~&~~$0.0514$~~&~~$0.0147$~~&~~$0.00457$~\\
\hline
$1P^{'}_{_1}$~~&~~$5853.65$~~&~~$0.811$~~&~~$0.747$~~&~~$0.796$~~&~~$1.03$~~&~~$0.163$~~&~~$0.0304$~~&~~$0.00652$~~&~~$0.00170$~\\
\hline
$2P^{'}_{_1}$~~&~~$6292.67$~~&~~$0.912$~~&~~$1.04$~~&~~$1.40$~~&~~$2.23$~~&~~$0.184$~~&~~$0.0423$~~&~~$0.0115$~~&~~$0.00369$~\\
 \hline
$3P^{'}_{_1}$~~&~~$6586.14$~~&~~$1.018$~~&~~$1.29$~~&~~$1.86$~~&~~$2.94$~~&~~$0.205$~~&~~$0.0523$~~&~~$0.0152$~~&~~$0.00485$~\\
\hline
\hline
\end{tabular}
\label{IV}
\end{table}

From Tables \ref{I}-\ref{V}, it can seen that the relations of $v^n$ are similar to that of $q^n$, and we will not repeat them, only provide some examples of $B^{(*)}_{(J)}$ mesons. Such as,
$v^2_{_{0^-}}(1S)\simeq v^2_{_{1^-}}(1S)\simeq 0.018<v^2_{_{1^+}}(1P)\simeq v^2_{_{2^{+}}}(1P)\simeq 0.024<v^2_{_{0^+}}(1P)\simeq v^2_{_{1^{+'}}}(1P)\simeq 0.027$.
These relations prove that our method is correct, because in the non-relativistic limit, the same wave function is used for the $0^-$ and $1^-$ states. While in our method, their wave functions are very different, but we obtain similar results for them. Similarly, in the HQET, $0^+$ and $1^{+'}$ belong to the $j=1/2$ states, and $1^+$ and $2^+$ states are the $j=3/2$ states. Therefore, without considering high-order corrections, the results of $0^+$ and $1^{+'}$ are the same, and $1^+$ and $2^+$ states share the same result. This is also consistent with our conclusion, while we have used completely different wave functions for them. It should be noted that although the wave function forms and normalization formulas for $1^+$ and $1^{+'}$ are the same, the calculated $v^n$ of them are different due to the different numerical values of their wave functions.

\begin{table}
\caption{ The average values of $q^n$ and $v^n$ of bottom quark inside $2^{+}$ states $B_2$, $B_{s2}$ and their excited states.}%, where the masses $M_{B^*_{_2}(5747)^0}=5737.20~\rm{MeV}$, $M_{B^*_{_{s2}}(5840)^0}=5839.86~\rm{MeV}$ are input.}
\begin{tabular}{cccccccccc}
\hline
\hline
State $(B_2)$~~&~~{Mass}~~&~~{$q$}~~&~~{$q^2$}~~&~~{$q^3$}~~&~~{$q^4$}~~&~~~{$v$}~~&~~{$v^2$}~~&~~{$v^3$}~~&~~{$v^4$}~~\\
\hline
\hline
$1^3P_{_2}$~~&~~$5737.20$~~&~~$0.728$~~&~~$0.589$~~&~~$0.528$~~&~~$0.524$~~&~~$0.146$~~&~~$0.0240$~~&~~$0.00433$~~&~~$0.000866$~~\\
\hline
$2^3P_{_2}$~~&~~$6215.21$~~&~~$0.873$~~&~~$0.912$~~&~~$1.07$~~&~~$1.36$~~&~~$0.176$~~&~~$0.0371$~~&~~$0.00878$~~&~~$0.00225$~~\\
\hline
$3^3P_{_2}$~~&~~$6499.78$~~&~~$0.958$~~&~~$1.11$~~&~~$1.44$~~&~~$2.04$~~&~~$0.193$~~&~~$0.0451$~~&~~$0.0118$~~&~~$0.00337$~~\\
\hline
\hline
State $(B_{s2})$~~&~~{Mass}~~&~~{$q$}~~&~~{$q^2$}~~&~~{$q^3$}~~&~~{$q^4$}~~&~~~{$v$}~~&~~{$v^2$}~~&~~{$v^3$}~~&~~{$v^4$}~~\\
\hline
\hline
$1^3P_{_2}$~~&~~$5839.86$~~&~~$0.775$~~&~~$0.672$~~&~~$0.647$~~&~~$0.689$~~&~~$0.156$~~&~~$0.0273$~~&~~$0.00530$~~&~~$0.00114$~~\\
\hline
$2^3P_{_2}$~~&~~$6300.50$~~&~~$0.902$~~&~~$0.991$~~&~~$1.23$~~&~~$1.66$~~&~~$0.182$~~&~~$0.0403$~~&~~$0.0101$~~&~~$0.00274$~~\\
 \hline
$3^3P_{_2}$~~&~~$6590.38$~~&~~$0.986$~~&~~$1.20$~~&~~$1.66$~~&~~$2.49$~~&~~$0.199$~~&~~$0.0489$~~&~~$0.0136$~~&~~$0.00411$~~\\
\hline
\hline
\end{tabular}
\label{V}
\end{table}

\subsection{The average values  $q^n$ and $v^n$ of a charm quark inside $D$ and $D_{s}$ mesons}

The average values $q^n$ and $v^n$ of a charm quark inside $0^-$ state $D$, $D_{s}$ and their excited states are shown in Tables \ref{VI}. From Table \ref{VI}, it can be seen that, for the average values $q^n$, we have the relations, $q^n_{_{D}}(1S)< q^n_{_{D}}(2S)< q^n_{_{D}}(3S)$ and $q^n_{_{D_s}}(1S)< q^n_{_{D_s}}(2S)< q^n_{_{D_s}}(3S)$ ($n=1,2,3,4;~m=1,2,3$). This means that, for $0^-$ state $D$, $D_{s}$, the $q^n$ in highly excited state is larger than in the ground state or lowly excited state. In the meantime, the values $q^n$ in the $D_{s}$ meson are bigger than the values in $D$ meson, $q^n_{_{D}}(mS)< q^n_{_{D_s}}(mS)$, this means that, the charm quark is bounded more deeply in $D$ meson than in $D_{s}$ meson. For average values $v^n$, we obtained relations similar to $q^n$, $v^n_{_{0^-}}(1S)<v_{_{0^-}}^n(2S)<v^n_{_{0^-}}(3S)$. Our results indicate that the speed expansion in the $D$ and $D_{s}$ mesons systems show poor convergence, especially for highly excited states, where the convergence is quite bad. This result also indicates that, for the $D$ and $D_{s}$ mesons, the relativistic corrections in highly excited state is greater than in ground and low excite states.
\begin{table}
\caption{ The average values of $q^n$ and $v^n$ of charm quark inside $0^{-}$ state $D$, $D_{s}$ and their excited states.}%, where the masses $M_{D^0,\bar{D}^0}=1864.84~\rm{MeV}$, $M_{D^\pm_s}=1968.35~\rm{MeV}$ are input.}
\begin{tabular}{cccccccccc}
\hline
\hline
State $(D)$~~&~~{Mass}~~&~~{$q$}~~&~~{$q^2$}~~&~~{$q^3$}~~&~~{$q^4$}~~&~~~{$v$}~~&~~{$v^2$}~~&~~{$v^3$}~~&~~{$v^4$}~~\\
\hline
\hline
$1^1S_{_0}$~~&~~$1864.84$~~&~~$0.509$~~&~~$0.317$~~&~~$0.238$~~&~~$0.220$~~&~~$0.314$~~&~~$0.121$~~&~~$0.0560$~~&~~$0.0319$~~\\
\hline
$2^1S_{_0}$~~&~~$2564.13$~~&~~$0.664$~~&~~$0.570$~~&~~$0.564$~~&~~$0.621$~~&~~$0.410$~~&~~$0.217$~~&~~$0.133$~~&~~$0.0902$~~\\
 \hline
$3^1S_{_0}$~~&~~$3041.83$~~&~~$0.795$~~&~~$0.780$~~&~~$0.866$~~&~~$1.05$~~&~~$0.491$~~&~~$0.297$~~&~~$0.204$~~&~~$0.153$~~\\
 \hline
 \hline
State $(D_{s})$~~&~~{Mass}~~&~~{$q$}~~&~~{$q^2$}~~&~~{$q^3$}~~&~~{$q^4$}~~&~~~{$v$}~~&~~{$v^2$}~~&~~{$v^3$}~~&~~{$v^4$}~~\\
\hline
\hline
$1^1S_{_0}$~~&~~$1968.35$~~&~~$0.549$~~&~~$0.369$~~&~~$0.299$~~&~~$0.296$~~&~~$0.339$~~&~~$0.140$~~&~~$0.0703$~~&~~$0.0430$~~\\
\hline
$2^1S_{_0}$~~&~~$2680.98$~~&~~$0.691$~~&~~$0.630$~~&~~$0.665$~~&~~$0.780$~~&~~$0.426$~~&~~$0.240$~~&~~$0.156$~~&~~$0.113$~~\\
 \hline
$3^1S_{_0}$~~&~~$3130.78$~~&~~$0.804$~~&~~$0.828$~~&~~$0.975$~~&~~$1.26$~~&~~$0.496$~~&~~$0.315$~~&~~$0.229$~~&~~$0.183$~~\\
\hline
\hline
\end{tabular}
\label{VI}
\end{table}

The average values $q^n$ and $v^n$ of a charm quark inside $1^-$ state $D^{*}$, $D^{*}_{s}$ and their excited states are shown in Tables \ref{VII}. From Tables \ref{VI} and Tables \ref{VII}, it can be seen that average values $q^n$ and $v^n$ inside $1^-$ state $D^{*}$, $D^{*}_{s}$ have similar relations to the cases of $0^-$ state $D$, $D_{s}$ mesons. In addition, the average values $q^n$ and $v^n$ inside $1^-$ state are not much different from the cases of $0^-$ state.
\begin{table}
\caption{ The average values of $q^n$ and $v^n$ of charm quark inside $1^{-}$ state $D^{*}$, $D^{*}_{s}$ and their excited states.}%, where the masses $M_{D^*(2007)^0}=2006.85~\rm{MeV}$, $M_{D^{*\pm}_s}=2112.20~\rm{MeV}$ are input.}
\begin{tabular}{cccccccccc}
\hline
\hline
State $(D^{*})$~~&~~{Mass}~~&~~{$q$}~~&~~{$q^2$}~~&~~{$q^3$}~~&~~{$q^4$}~~&~~~{$v$}~~&~~{$v^2$}~~&~~{$v^3$}~~&~~{$v^4$}~~\\
\hline
\hline
$1^3S_{_1}$~~&~~$2006.85$~~&~~$0.533$~~&~~$0.342$~~&~~$0.260$~~&~~$0.236$~~&~~$0.329$~~&~~$0.130$~~&~~$0.0613$~~&~~$0.0343$~~\\
\hline
$2^3S_{_1}$~~&~~$2627.40$~~&~~$0.677$~~&~~$0.589$~~&~~$0.585$~~&~~$0.642$~~&~~$0.418$~~&~~$0.224$~~&~~$0.138$~~&~~$0.0932$~~\\
\hline
$3^3S_{_1}$~~&~~$3115.95$~~&~~$0.783$~~&~~$0.716$~~&~~$0.729$~~&~~$0.801$~~&~~$0.483$~~&~~$0.273$~~&~~$0.171$~~&~~$0.116$~~\\
\hline
\hline
State $(D^{*}_{s})$~~&~~{Mass}~~&~~{$q$}~~&~~{$q^2$}~~&~~{$q^3$}~~&~~{$q^4$}~~&~~~{$v$}~~&~~{$v^2$}~~&~~{$v^3$}~~&~~{$v^4$}~~\\
\hline
\hline
$1^3S_{_1}$~~&~~$2112.20$~~&~~$0.572$~~&~~$0.394$~~&~~$0.323$~~&~~$0.314$~~&~~$0.352$~~&~~$0.150$~~&~~$0.0760$~~&~~$0.0456$~~\\
\hline
$2^3S_{_1}$~~&~~$2719.85$~~&~~$0.712$~~&~~$0.658$~~&~~$0.695$~~&~~$0.807$~~&~~$0.439$~~&~~$0.251$~~&~~$0.163$~~&~~$0.117$~~\\
 \hline
$3^3S_{_1}$~~&~~$3200.17$~~&~~$0.833$~~&~~$0.878$~~&~~$1.05$~~&~~$1.37$~~&~~$0.514$~~&~~$0.334$~~&~~$0.247$~~&~~$0.198$~~\\
\hline
\hline
\end{tabular}
\label{VII}
\end{table}

Furthermore, compared $v^n$ of the bottom quark inside the $B$ and $B_{s}$ mesons (in Tables \ref{I}), it is found that, for the same state, the average values $v^n$ of the bottom quark inside the $B$ and $B_{s}$ mesons are much smaller than the charm quark inside $D$ and $D_{s}$ mesons. For instance, $v^2_{_{B(1S)}}=0.0185\ll v^2_{_{D(1S)}}=0.121$ and $v^2_{_{B_{s}(2S)}}=0.0356\ll v^2_{_{D_{s}(2S)}}=0.240$, in $0^-$ state. This is because that, the bottom quark is more heavy than charm, it has a small velocity in heavy-light meson, so it's speed expansion behavior is better than the charm quark case. This indicate that there are smaller relativistic corrections in the $B$ and $B_{s}$ mesons than in the $D$ and $D_{s}$ mesons. For the $1^-$ state $B^{*}$, $B^{*}_{s}$ and $D^{*}$, $D^{*}_{s}$, the same properties also exist. Refs. \cite{GuoLiWang2020,W.L.Sang2015} give a similar conclusion about quarkonia.

The average values $q^n$ and $v^n$ of a charm quark inside $0^+$, $1^+$($P_{_1}$, $P^{'}_{_1}$) and $2^+$ $D_J$ and $D_{sJ}$ mesons and their excited states are shown in Tables \ref{VIII}, \ref{IX} and \ref{X}, respectively. For average values $q^n$ of these $P$ wave $D_J(nP)$ and $D_{sJ}(nP)$, we also have the relations, $q^n_{_{D_J}}(1P)< q^n_{_{D_J}}(2P)< q^n_{_{D_J}}(3P)$, $q^n_{_{D_{sJ}}}(1P)< q^n_{_{D_{sJ}}}(2P)< q^n_{_{D_{sJ}}}(3P)$, and $q^n_{_{D_J}}(mP)< q^n_{_{D_{sJ}}}(mP)$.
Compared with the results of $S$ waves, the relations $q^n_{_{D_J}}(mS)< q^n_{_{D_J}}(mP)$ and $q^n_{_{D_{sJ}}}(mS)< q^n_{_{D_{sJ}}}(mP)$ also exist. For the average values $v^n$, it have the similar relations to $q^n$. In addition, the $v^n_{_{B_{J},B_{sJ}}}(mP)$ of bottom quark inside the $B_J$ and $B_{sJ}$ mesons are much smaller than the corresponding $v^n_{_{D_{J},D_{sJ}}}(mP)$ of charm quark inside the $D_J$ and $D_{sJ}$ mesons, which indicate there are much bigger relativistic corrections in the $D_J$ and $D_{sJ}$ meson systems than the $B_J$ and $B_{sJ}$ meson systems for same $J^{P}$ state.

\begin{table}
\caption{ The average values of $q^n$ and $v^n$ of charm quark inside $0^{+}$ state $D^{*}_{0}$, $D^{*}_{s0}$ and their excited states.}%, where the masses $M_{D^*_0(2300)^0}=2300.00~\rm{MeV}$, $M_{D^*_{s0}(2317)^\pm}=2317.80~\rm{MeV}$ are input.}
\begin{tabular}{cccccccccc}
\hline
\hline
State $(D^{*})$~~&~~{Mass}~~&~~{$q$}~~&~~{$q^2$}~~&~~{$q^3$}~~&~~{$q^4$}~~&~~~{$v$}~~&~~{$v^2$}~~&~~{$v^3$}~~&~~{$v^4$}~~\\
\hline
\hline
$1^3P_{_0}$~~&~~$2300.00$~~&~~$0.617$~~&~~$0.426$~~&~~$0.330$~~&~~$0.289$~~&~~$0.381$~~&~~$0.162$~~&~~$0.0775$~~&~~$0.0419$~~\\
\hline
$2^3P_{_0}$~~&~~$2914.74$~~&~~$0.766$~~&~~$0.697$~~&~~$0.712$~~&~~$0.795$~~&~~$0.473$~~&~~$0.266$~~&~~$0.167$~~&~~$0.115$~~\\
 \hline
$3^3P_{_0}$~~&~~$3365.75$~~&~~$0.873$~~&~~$0.908$~~&~~$1.04$~~&~~$1.26$~~&~~$0.539$~~&~~$0.346$~~&~~$0.242$~~&~~$0.183$~~\\
\hline
\hline
State $(D^{*}_{s})$~~&~~{Mass}~~&~~{$q$}~~&~~{$q^2$}~~&~~{$q^3$}~~&~~{$q^4$}~~&~~~{$v$}~~&~~{$v^2$}~~&~~{$v^3$}~~&~~{$v^4$}~~\\
\hline
\hline
$1^3P_{_0}$~~&~~$2317.80$~~&~~$0.637$~~&~~$0.457$~~&~~$0.367$~~&~~$0.334$~~&~~$0.393$~~&~~$0.174$~~&~~$0.0864$~~&~~$0.0485$~~\\
\hline
$2^3P_{_0}$~~&~~$2995.27$~~&~~$0.781$~~&~~$0.741$~~&~~$0.797$~~&~~$0.937$~~&~~$0.482$~~&~~$0.282$~~&~~$0.187$~~&~~$0.136$~~\\
 \hline
$3^3P_{_0}$~~&~~$3339.07$~~&~~$0.886$~~&~~$0.924$~~&~~$1.10$~~&~~$1.42$~~&~~$0.547$~~&~~$0.352$~~&~~$0.259$~~&~~$0.207$~~\\
\hline
\hline
\end{tabular}
\label{VIII}
\end{table}
\begin{table}
\caption{ The average values of $q^n$ and $v^n$ of charm quark inside $1^{+}$ state $D_{1}(P_{1})$, $D^{'}_{1}(P^{'}_{1})$, $D_{s1}(P_{1})$ and $D^{'}_{s1}(P^{'}_{1})$ and their excited states.}%, where the masses $M_{D_1(2430)^0}=2412.00~\rm{MeV}$, $M_{D_{1}(2420)^\pm}=2317.80~\rm{MeV}$, $M_{D_{s1}(2460)^\pm}=2459.50~\rm{MeV}$, $M_{D_{s1}(2536)^\pm}=2535.11~\rm{MeV}$ are input.}
\begin{tabular}{cccccccccc}
\hline
\hline
State $(D^{(')})$~~&~~{Mass}~~&~~{$q$}~~&~~{$q^2$}~~&~~{$q^3$}~~&~~{$q^4$}~~&~~~{$v$}~~&~~{$v^2$}~~&~~{$v^3$}~~&~~{$v^4$}~~\\
\hline
\hline
$1P_{_1}$~~&~~$2412.00$~~&~~$0.639$~~&~~$0.452$~~&~~$0.351$~~&~~$0.301$~~&~~$0.395$~~&~~$0.172$~~&~~$0.0827$~~&~~$0.0437$~~\\
\hline
$2P_{_1}$~~&~~$2919.01$~~&~~$0.773$~~&~~$0.701$~~&~~$0.707$~~&~~$0.767$~~&~~$0.477$~~&~~$0.267$~~&~~$0.166$~~&~~$0.111$~~\\
 \hline
$3P_{_1}$~~&~~$3259.88$~~&~~$0.880$~~&~~$0.910$~~&~~$1.02$~~&~~$1.23$~~&~~$0.543$~~&~~$0.347$~~&~~$0.241$~~&~~$0.178$~~\\
\hline
$1P^{'}_{_1}$~~&~~$2422.00$~~&~~$0.652$~~&~~$0.472$~~&~~$0.380$~~&~~$0.343$~~&~~$0.402$~~&~~$0.180$~~&~~$0.0894$~~&~~$0.0498$~~\\
\hline
$2P^{'}_{_1}$~~&~~$2949.37$~~&~~$0.779$~~&~~$0.716$~~&~~$0.736$~~&~~$0.822$~~&~~$0.481$~~&~~$0.273$~~&~~$0.173$~~&~~$0.119$~~\\
 \hline
$3P^{'}_{_1}$~~&~~$3278.60$~~&~~$0.882$~~&~~$0.920$~~&~~$1.04$~~&~~$1.27$~~&~~$0.545$~~&~~$0.351$~~&~~$0.245$~~&~~$0.185$~~\\
\hline
\hline
State $(D^{(')}_{s})$~~&~~{Mass}~~&~~{$q$}~~&~~{$q^2$}~~&~~{$q^3$}~~&~~{$q^4$}~~&~~~{$v$}~~&~~{$v^2$}~~&~~{$v^3$}~~&~~{$v^4$}~~\\
\hline
\hline
$1P_{_1}$~~&~~$2459.00$~~&~~$0.661$~~&~~$0.487$~~&~~$0.396$~~&~~$0.357$~~&~~$0.408$~~&~~$0.185$~~&~~$0.0932$~~&~~$0.0519$~~\\
\hline
$2P_{_1}$~~&~~$3015.69$~~&~~$0.787$~~&~~$0.746$~~&~~$0.793$~~&~~$0.911$~~&~~$0.486$~~&~~$0.284$~~&~~$0.186$~~&~~$0.132$~~\\
 \hline
$3P_{_1}$~~&~~$3356.12$~~&~~$0.888$~~&~~$0.933$~~&~~$1.10$~~&~~$1.41$~~&~~$0.548$~~&~~$0.356$~~&~~$0.260$~~&~~$0.204$~~\\
\hline
$1P^{'}_{_1}$~~&~~$2535.11$~~&~~$0.688$~~&~~$0.528$~~&~~$0.451$~~&~~$0.433$~~&~~$0.424$~~&~~$0.201$~~&~~$0.106$~~&~~$0.0629$~~\\
\hline
$2P^{'}_{_1}$~~&~~$3034.09$~~&~~$0.791$~~&~~$0.757$~~&~~$0.817$~~&~~$0.960$~~&~~$0.489$~~&~~$0.288$~~&~~$0.192$~~&~~$0.139$~~\\
 \hline
$3P^{'}_{_1}$~~&~~$3368.47$~~&~~$0.894$~~&~~$0.940$~~&~~$1.12$~~&~~$1.45$~~&~~$0.551$~~&~~$0.358$~~&~~$0.264$~~&~~$0.211$~~\\
\hline
\hline
\end{tabular}
\label{IX}
\end{table}
\begin{table}
\caption{ The average values of $q^n$ and $v^n$ of charm quark inside $2^{+}$ state $D_{2}$, $D_{s2}$ and their excited states.}% where the masses $M_{D^*_2(2460)^0}=2461.10~\rm{MeV}$, $M_{D^*_{s2}(2573)^\pm}=2569.10~\rm{MeV}$ are input.}
\begin{tabular}{cccccccccc}
\hline
\hline
State $(D_2)$~~&~~{Mass}~~&~~{$q$}~~&~~{$q^2$}~~&~~{$q^3$}~~&~~{$q^4$}~~&~~~{$v$}~~&~~{$v^2$}~~&~~{$v^3$}~~&~~{$v^4$}~~\\
\hline
\hline
$1^3P_{_2}$~~&~~$2461.10$~~&~~$0.650$~~&~~$0.465$~~&~~$0.364$~~&~~$0.311$~~&~~$0.401$~~&~~$0.177$~~&~~$0.0857$~~&~~$0.0452$~~\\
\hline
$2^3P_{_2}$~~&~~$2959.65$~~&~~$0.785$~~&~~$0.719$~~&~~$0.728$~~&~~$0.789$~~&~~$0.485$~~&~~$0.274$~~&~~$0.171$~~&~~$0.114$~~\\
 \hline
$3^3P_{_2}$~~&~~$3329.47$~~&~~$0.900$~~&~~$0.968$~~&~~$1.12$~~&~~$1.36$~~&~~$0.556$~~&~~$0.369$~~&~~$0.264$~~&~~$0.197$~~\\
 \hline
 \hline
State $(D_{s2})$~~&~~{Mass}~~&~~{$q$}~~&~~{$q^2$}~~&~~{$q^3$}~~&~~{$q^4$}~~&~~~{$v$}~~&~~{$v^2$}~~&~~{$v^3$}~~&~~{$v^4$}~~\\
\hline
\hline
$1^3P_{_2}$~~&~~$2569.10$~~&~~$0.683$~~&~~$0.517$~~&~~$0.429$~~&~~$0.390$~~&~~$0.421$~~&~~$0.197$~~&~~$0.101$~~&~~$0.0567$~~\\
\hline
$2^3P_{_2}$~~&~~$3050.19$~~&~~$0.794$~~&~~$0.754$~~&~~$0.799$~~&~~$0.911$~~&~~$0.490$~~&~~$0.287$~~&~~$0.188$~~&~~$0.132$~~\\
 \hline
$3^3P_{_2}$~~&~~$3418.45$~~&~~$0.929$~~&~~$0.979$~~&~~$1.15$~~&~~$1.47$~~&~~$0.574$~~&~~$0.373$~~&~~$0.271$~~&~~$0.214$~~\\
\hline
\hline
\end{tabular}
\label{X}
\end{table}

\section{Summary}

In summary, based on the instantaneous Bethe-Salpeter equation method, we calculate the average values $\overline{|\vec{q}|^n}\equiv q^n$, including the kinetic energy $\mu^2_{_\pi}$ (or $-\lambda_{_1}$), and speed powers $\overline{|\vec{v}|^n} \equiv v^n$ ($n=1,2,3,4$) of a heavy quark inside $S$ wave and $P$ wave heavy-light mesons. We obtain $\mu^2_{_\pi}=0.455$ GeV$^2$ for the $B$ meson, which is consistent with the experimental result $0.464\pm 0.076$ GeV$^2$. For the $B_{s}$, $D$ and $D_{s}$ mesons, the $\mu^2_{_\pi}$ are $0.530$ GeV$^2$, $0.317$ GeV$^2$ and $0.369$ GeV$^2$, respectively. So, there is a clear relationship as follows: $q^2_{_{B}}(nS)< q^2_{_{B_s}}(nS)$, and $q^2_{_{D}}(nS)< q^2_{_{D_s}}(nS)$.
We also find the following relationships  $q^n_{_{0^-}}(mS)\approx q^n_{_{1^-}}(mS)$,  $q^n_{_{0^+}}(mP)\approx q^n_{_{1^{+'}}}(mP^{'})> q^n_{_{1^+}}(mP)\approx q^n_{_{2^+}}(mP)$, and $q^n(mS)< q^n(mP)$, etc., for the same constituent heavy-light mesons.
As expected values, for example, $\overline{|\vec{q}|^4}\neq\overline{|\vec{q}|^2}^2$. If we replace momentum with speed, we have the similar relationships.

{\bf Acknowledgments}
This work was supported in part by the National Natural Science Foundation of China (NSFC) under the Grants Nos. 12075073, 12375085, 12075301, and 12075074, the Natural Science Foundation of Hebei province under the Grant No. A2021201009, Post-graduate's Innovation Fund Project of Hebei University under the Grant No. HBU2022BS002.

\section{APPENDIX}
\setcounter{equation}{0}
\renewcommand\theequation{A.\arabic{equation}} %£¨A1£©
\subsection{Introduction of The Bethe-Salpeter Equation and Salpeter Equation}

The Bethe-Salpeter (BS) equation which is used to describe the heavy-light meson can be written as \cite{E.E.SandH.A.B.1951}
\begin{eqnarray}\label{BSE}
(\slashed{p}_{_Q}-m_{_Q})\chi_{_P}(P,q)(\slashed{p}_{_q}+m_{_q})=i\int\frac{d^4k}{(2\pi)^{4}}V(P,k,q)\chi_{_P}(k),
\end{eqnarray}
where, $\chi_{_P}(q)$ is the relativistic BS wave function, $V(P,k,q)$ is the interaction kernel between quark and antiquark. $P$ and $q(k)$ are the total momentum and the relative momentum of the meson, respectively. $p_{_Q}$ and $p_{_q}$ are the momenta of the heavy quark and light antiquark, $m_{_Q}$ and $m_{_q}$ are the constituent masses of the heavy quark and light antiquark, respectively. The meson momentum $P$ and relative momentum $q$ have related by, $$p_{_Q}=\alpha_{_1}P+q,~~~~~\alpha_{_1}=\frac{m_{_Q}}{m_{_Q}+m_{_q}},~~~~~p_{_q}=\alpha_{_2}P-q,~~~~~\alpha_{_2}=\frac{m_{_q}}{m_{_Q}+m_{_q}}.$$ For momentum decomposite, $q^{\mu}=q^{\mu}_{_\parallel}+q^{\mu}_{_\perp}$, $q^{\mu}_{_\parallel}\equiv\frac{P\cdot q}{M^2}P^{\mu},~q^{\mu}_{_\perp}\equiv q^{\mu}-q^{\mu}_{_\parallel}.$ Correspondingly, we have two Lorentz invariant variables, $q_{_P}=\frac{P\cdot q}{M},~q_{_T}=\sqrt{q^2_{_P}-q^2}=\sqrt{-q^2_{_\perp}}.$ In the rest frame of the bound state, $i.e.$, $\vec{P}=0$, they turn to the usual components $q^0$ and $|\vec{q}|.$

The Salpeter equation is instantaneous version of the BS equation. In the instantaneous approach and in the rest frame of the meson, $P=(M,\textbf{0})$, and the interaction kernel $V(P,k,q)$ becomes to $V(k_{_\perp},q_{_\perp})$. After integrating over $q_{_0}$, then the BS wave function $\chi_{_P}(P,q)$ becomes to three dimensional BS wave function $\varphi(q_{_\perp})$,
\begin{eqnarray}\label{SE}
\varphi_p(q^{\mu}_{_\perp})\equiv i\int\frac{dq_{_p}}{(2\pi)}\chi_{_P}(q^{\mu}_{_{_\parallel}},q^{\mu}_{_\perp}),
\end{eqnarray}
and new integration kernel
\begin{eqnarray}\label{ik}
\eta(q^{\mu}_{_\perp})\equiv\int\frac{k^2_{_T}dk_{_T}ds}{(2\pi)^2}V(k_{_\perp},q_{_{\perp}})\varphi_{P}(k^{\mu}_{_\perp}).\nonumber
\end{eqnarray}

The useful notations
$$\omega_{_i}=\sqrt{m^2_{_i}+q^2_{_T}},~~~~~
\Lambda^{\pm}_{_i}(q_{_\perp})=\frac{1}{2\omega_{_i}}\big[\frac{\slashed{P}}{M}\omega_{_i}\pm J_{_i}(m_{_i}+\slashed{q}_{_\perp})\big],$$
$$\varphi^{\pm\pm}_{_P}(q_{_\perp})=\Lambda^{\pm}_{_1}(q_{_\perp})\frac{\slashed{P}}{M}\varphi_{_P}(q_{_\perp})
\frac{\slashed{P}}{M}\Lambda^{\pm}_{_2}(q_{_\perp}).$$
where $i=1,2$ for heavy quark and light antiquark, that is, $\omega_{_1} = \omega_{_Q}$, $\omega_{_2} = \omega_{_q}$, and $J(i)=(-1)^{i+1}$.
Then, BS equation Eq. (\ref{BSE}) can be written as
\begin{eqnarray}\label{SE22}
\chi(q)=S(p_{_Q})\eta(q_{_\perp})S(-p_{_q}).
\end{eqnarray}
$S(p_{_Q})$ and $S(-p_{_q})$ represent quark and anti-quark propagators, respectively. Its time components can be separated and decomposed into the following form
\begin{eqnarray}
 S(p_{_Q})=\frac{i\Lambda^{+}_{_1}}{q_{_P}+\alpha_{_1}M-\omega_{_Q}+i\epsilon}+\frac{i\Lambda^{-}_{_1}}{q_{_P}+
 \alpha_{_1}M+\omega_{_Q}-i\epsilon},\nonumber\\
 S(-p_{_q})=\frac{i\Lambda^{+}_{_2}}{q_{_P}-\alpha_{_2}M+\omega_{_q}-i\epsilon}+\frac{i\Lambda^{-}_{_2}}{q_{_P}+
 \alpha_{_2}M-\omega_{_q}+i\epsilon}.
\end{eqnarray}
 Then, we take the new integration kernel and integrate over $q_{_0}$ in Eq. (\ref{SE22}) on both sides, we get the Salpeter equation,
\begin{eqnarray}
\varphi(q_{_\perp})=\frac{\Lambda^{+}_{_1}(q_{_\perp})\eta(q_{_\perp})\Lambda^{+}_{_2}(q_{_\perp})}{M-\omega_{_Q}-\omega_{_q}}-
\frac{\Lambda^{-}_{_1}(q_{_\perp})\eta(q_{_\perp})\Lambda^{-}_{_2}(q_{_\perp})}{M+\omega_{_Q}+\omega_{_q}}.
\end{eqnarray}

Using the notions $\varphi^{\pm\pm}(q_{_\perp})$, the Salpeter wave function can be separated into four terms
\begin{eqnarray}
\varphi_{_P}(q_{_\perp})=\varphi^{++}_{_P}(q_{_\perp})+\varphi^{+-}_{_P}(q_{_\perp})+\varphi^{-+}_{_P}(q_{_\perp})
+\varphi^{--}_{_P}(q_{_\perp}),
\end{eqnarray}
where, $\varphi^{++}_{_P}$ is the positive wave function, $\varphi^{--}_{_P}$ is the negative wave function. Using the relations of projection operators, then obtain the so-called Salpeter equation, $i.e.$, a set of coupled equations: \cite{E.E. S1952}
\begin{eqnarray}\label{SE2}
&&(M-\omega_Q-\omega_q)\varphi^{++}_{_P}(q_{_\perp})=\Lambda^{+}_1(q_{_\perp})\eta_{_P}(q_{_\perp})\Lambda^{+}_2(q_{_\perp}),\nonumber\\
&&(M+\omega_Q+\omega_q)\varphi^{--}_{_P}(q_{_\perp})=\Lambda^{-}_1(q_{_\perp})\eta_{_P}(q_{_\perp})\Lambda^{-}_2(q_{_\perp}),\nonumber\\
&&\hspace{1.3cm}\varphi^{+-}_{_P}(q_{_\perp})=\varphi^{-+}_{_P}(q_{_\perp})=0.
\end{eqnarray}

In our model, Cornell potential is chosen as the instantaneous interaction kernel
\begin{eqnarray}
V=V_0+V_{s}(r)+\gamma_{_0}\otimes\gamma^{0}V_{v}(r)=V_{0}+\lambda r-\gamma_{_0}\otimes\gamma^{0}\frac{4}{3}\frac{\alpha_{_s}}{r},
\end{eqnarray}
the running coupling constant $\alpha_{_s}(\vec{q})=\frac{12\pi}{33-2N_{_f}}\frac{1}{log(a+\frac{\vec{q}^2}{\Lambda_{_{QCD}}})}.$ To avoid infrared divergence, a factor $e^{-\alpha r}$ is introduced. Then, in momentum space
\begin{eqnarray}\label{CP}
&&\hspace{2.3cm}V(\vec{q})=(2\pi)^3V_{s}(\vec{q})+\gamma_{_0}\otimes\gamma^{0}V_{v}(\vec{q}),\nonumber\\
&&V_{s}(\vec{q})=-(\frac{\lambda}{\alpha}+V_{_0})\delta^3(\vec{q})+\frac{\lambda}{\pi^2}\frac{1}{(\vec{q}^2+\alpha^2)^2},
~~~V_{v}(\vec{q})=-\frac{2}{3\pi^2}\frac{\alpha_{_s}(\vec{q})}{\vec{q}^2+\alpha^2}.
\end{eqnarray}


\begin{thebibliography}{50}
\vspace{3mm}
\bibitem{BigiI1994}Karos I. Y. Bigi, Mikhail A. Shifman, N. G. Uraltsev, A. I. Vainshtein, Int. J. Mod. Phys. A 9, 2467 (1994).
\bibitem{FalkA1993}Adam F. Falk, Matthias Neubert, Phys. Rev. D 47, 2965 (1993).
\bibitem{NeubertM2005}Neubert M, Phys. Rev. D 72, 074025 (2005).
\bibitem{N.Isgur1989}N. Isgur and M. B. Wise, Phys. Lett. B 232, 113 (1989).
\bibitem{N.Isgur1990}N. Isgur and M. B. Wise, Phys. Lett. B 237, 527 (1990).
\bibitem{H.Georgi1991}H. Georgi, Phys. Lett. B 264, 447 (1991).
\bibitem{M.Neubert1994}M. Neubert, Phys. Rep. 245, 259 (1994).
\bibitem{O.L.Buchm2006}O. L. Buchmuller and H. U. Flacher, Phys. Rev. D 73, 073008 (2006).
\bibitem{BriereRA2002}Roy A. Briere $et~al$., (CLEO Collaboration), CLEO-CONF 02-10 (2002), hep-ex/0209024.
\bibitem{ChenS2001}S. Chen $et~al$., (CLEO Collaboration), Phys. Rev. Lett. 87, 251807 (2001).
\bibitem{CroninHennessyD2001}D. Cronin-Hennessy $et~al$., (CLEO Collaboration), Phys. Rev. Lett. 87, 251808 (2001).
\bibitem{HwangDS1997}Dae Sung Hwang, C. S. Kim, Wuk Namgung, Phys. Lett. B 406, 117 (1997).
\bibitem{FazioFD1996}Fulvia De Fazio, Mod. Phys. Lett. A 11, 2693 (1996).
\bibitem{KimCS2004}C. S. Kim and Guo-Li Wang, Phys. Lett. B 584, 285 (2004).
\bibitem{Caswell1986}W. E. Caswell and G. P. Lepage, Phys. Lett. B 167, 437 (1986).
\bibitem{Bodwin1997}Geoffrey T. Bodwin, Eric Braaten, G. Peter Lepage, Phys. Rev. D 51, 1125 (1995); 55, 5853(E) (1997).
\bibitem{GeoffreyT2006}Geoffrey T. Bodwin, Daekyoung Kang, Jungil Lee, Phys. Rev. D 74, 014014 (2006).
\bibitem{G.T.Bodwin2004}Geoffrey T. Bodwin and Jungil Lee, Phys. Rev. D 69, 054003 (2004).
\bibitem{R.L.Zhu2017}Ruilin Zhu, Yan Ma, Xin-Ling Han, Zhen-Jun Xiao, Phys. Rev. D 95, 094012 (2017).
\bibitem{F.Feng2017}Feng Feng, Yu Jia, Wen-Long Sang, Phys. Rev. Lett. 119, 252001 (2017).
\bibitem{W.Wang2017}Wei Wang, Ji Xu, Deshan Yang, Shuai Zhao, J. High Energy Phys. 12,  012 (2017).
\bibitem{D.Ebert2006}D. Ebert and A. P. Martynenko, Phys. Rev. D 74, 054008 (2006).
\bibitem{D.Ebert2009}D. Ebert, R. N. Faustov, V. O. Galkin, and A. P. Martynenko, Phys. Lett. B 672, 264 (2009).
\bibitem{G.T.Bodwin2002}Geoffrey T. Bodwin, D. K. Sinclair, S. Kim, Phys. Rev. D 65, 054504 (2002).
\bibitem{Z.K.Geng2019}Zi-Kan Geng, Tianhong Wang, Yue Jiang, Geng Li, Xiao-Ze Tan, Guo-Li Wang, Phys. Rev. D 99, 013006 (2019).
\bibitem{G.L.Wang2020}Guo-Li Wang and Xing-Gang Wu, Chin. Phys. C 44, 063104 (2020).
\bibitem{W.Li2023}Wei Li, Su-Yan Pei, Tianhong Wang, Ying-Long Wang, Tai-Fu Feng, Guo-Li Wang, Phys. Rev. D 107, 113002 (2023).
\bibitem{M.Gremm1997}Martin Gremm and Anton Kapustin, Phys. Lett. B 407, 323 (1997).
\bibitem{R.L.Zhu2018}Ruilin Zhu, Nucl. Phys. B 931, 359 (2018).
\bibitem{W.Buchmuller1981}W. Buchmuller and S. H. H. Tye, Phys. Rev. D 24, 132 (1981).
\bibitem{V.V.Braguta2009}V. V. Braguta, A. K. Likhoded, A. V. Luchinsky, Phys. Rev. D 79, 074004 (2009).
\bibitem{C.W.Hwang2009}Chien-Wen Hwang, J. High Energy Phys. 10, 074 (2009).
\bibitem{H.K.Guo2011}Huai-Ke Guo, Yan-Qing Ma, Kuang-Ta Chao, Phys. Rev. D 83, 114038 (2011).
\bibitem{GuoLiWang2020}Guo-Li Wang, Tai-Fu Feng, Xing-Gang Wu, Phys. Rev. D 101, 116011 (2020).
\bibitem{E.E. S1952}E. E. Salpeter, Phys. Rev. 87, 328 (1952).
\bibitem{E.E.SandH.A.B.1951}E. E. Salpeter and H. A. Bethe, Phys. Rev. 84, 1232 (1951).
%\bibitem{G.L.Wang2008}Guo-Li Wang, Int. J. Mod. Phys. A 23, 3263 (2008).
\bibitem{C.Hsi.Chang2010}Chao-Hsi Chang, Guo-Li Wang, Sci. China. Phys. Mech. Astron. 53, 2005 (2010).
\bibitem{G.L.Wang2022}Guo-Li Wang, Tianghong Wang, Qiang Li, Chao-Hsi Chang, J. High Energy Phys. 05, 006 (2022).
\bibitem{S.C.Li2018}Si-Chen Li, Tianhong Wang, Yue Jiang, Xiao-Ze Tan, Qiang Li, Guo-Li Wang, Chao-Hsi Chang, Phys. Rev. D 97, 5 (2018).
\bibitem{T.h.Wang2017}Tianhong Wang, Zhi-Hui Wang, Yue Jiang, Li-Bo Jiang, Guo-Li Wang, Eur. Phys. J. C 77, 1 (2017).
\bibitem{Z.H.Wang2022}Zhi-Hui Wang and Guo-Li Wang, Phys. Rev. D 106, 5 (2022).
\bibitem{H.F.Fu2012}Zhi-Hui Wang, Guo-Li Wang, Hui-Feng Fu, Yue Jiang, Int. J. Mod. Phys. A 27, 1250027 (2012).
\bibitem{G.-L.Wang2006}Guo-Li Wang, Phys. Lett. B 633, 492 (2006).
\bibitem{G.-L.Wang2007}Guo-Li Wang, Phys. Lett. B 650, 15 (2007).
\bibitem{Q.Li2020}Qiang Li, Tianhong Wang, Yue Jiang, Guo-Li Wang, Chao-Hsi Chang, Phys. Rev. D 100, 076020 (2020).
\bibitem{G.-L.Wang2009}Guo-Li Wang, Phys. Lett. B 674, 172 (2009).
%\bibitem{G.Altarelli1982}Guido Altarelli, N. Cabibbo, G. Corbo, L. Maiani, G. Martinelli, Nucl. Phys. B 208, 365 (1982).
\bibitem{E.Eichten1978}E. Eichten, K. Gottfried, T. Kinoshita, K. D. Lane, Tung-Mow Yan, Phys. Rev. D 17, 3090 (1978). [Erratum ibid. D 21, 313 (1980)].
\bibitem{S.G1985}S. Godfrey and Nathan Isgur, Phys. Rev. D 32, 189 (1985).
\bibitem{K.A.Olive2022}K. A. Olive $et~al$., (Partile Data Group), Prog. Theor. Exp. Phys. 2022, 083C01 (2022).
\bibitem{Yasmine2023}Yasmine Sara Amhis $et~al$., (HFLAV Collaboration), Phys. Rev. D 107, 052008 (2023).
\bibitem{Belle2021}R. van Tonder $et~ al$., (Belle collaboration), Phys. Rev. D 104, 112011 (2021).
\bibitem{Belle-II2022}F. Abudinen $et ~al$., (Belle II collaboration), Phys. Rev. D 107, 072002 (2023).
\bibitem{I.I.Bigi1995}Ikaros I. Y. Bigi, Mikhail A. Shifman, N. G. Uraltsev, Arkady I. Vainshtein, Phys. Rev. D 52, 196 (1995).
\bibitem{M.Shifman1995}Mikhail A. Shifman, N. G. Uraltsev, Arkady I. Vainshtein, Phys. Rev. D 51, 2217 (1995).
\bibitem{M.Voloshin1995}M. B. Voloshin, Surveys High Energ. Phys. 8, 2751 (1995).
\bibitem{E.Bagan1996}E. Bagan, Patricia Ball, Vladimir M. Braun, P. Gosdzinsky, Phys. Lett. B 342, 362 (1995).
\bibitem{Patricia1994}Patricia Ball and Vladimir M. Braun, Phys. Rev. D 49, 9307291 (1994).
\bibitem{Nikolai1998}Nikolai Uraltsev, Proc. Int. Sch. Phys. Fermi 137,  329 (1998).
\bibitem{P.Gambino2011}Paolo Gambino, J. High Energy Phys. 09, 055 (2011).
\bibitem{A.Alberti2015}Andrea Alberti, Paolo Gambino, Kristopher J. Healey, Soumitra Nandi, Phys. Rev. Lett., 114, 061802 (2015).
\bibitem{M.Bordone2021}Marzia Bordone, Bernat Capdevila, Paolo Gambino, Phys. Lett. B 822, 136679 (2021).
\bibitem{W.L.Sang2015}Wen-Long Sang, Feng Feng, Yu-Qi Chen, Phys. Rev. D 92, 014025 (2015).
\end{thebibliography}
\end{document}